\documentclass[sigconf,screen,review=false,natbib]{acmart}
\AtBeginDocument{%
\providecommand\BibTeX{{%
\normalfont B\kern-0.5em{\scshape i\kern-0.25em b}\kern-0.8em\TeX}}}

\ccsdesc[500]{Theory of computation~Automated reasoning}

\copyrightyear{2023}
\acmYear{2023}
\setcopyright{rightsretained}
\acmConference[AICCC 2023]{2023 6th Artificial Intelligence and Cloud Computing Conference (AICCC)}{December 16--18, 2023}{Kyoto, Japan} \acmBooktitle{2023 6th Artificial Intelligence and Cloud Computing Conference (AICCC) (AICCC 2023), December 16--18, 2023, Kyoto, Japan}\acmDOI{10.1145/3639592.3639622}
\acmISBN{979-8-4007-1622-5/23/12}

\usepackage{amsmath}
\usepackage{multirow}
\usepackage{microtype}
\usepackage{tikz,tikz-qtree}
\usetikzlibrary{positioning}
\usetikzlibrary{arrows.meta, shapes.misc, positioning}
\tikzstyle{block}=[draw,fill=white,rectangle]
\tikzstyle{every node}=[font=\small]
\usepackage{subfig}
\theoremstyle{definition}

\usepackage[ruled, vlined, linesnumbered]{algorithm2e}
\SetKw{True}{true}
\SetKw{False}{false}
\SetKwInOut{Output}{Output}
\SetKwInOut{Input}{Input}
\SetKwData{typeInt}{Int}
\SetKwData{typeRat}{Rat}
\SetKwData{Defined}{Defined}
\SetKwProg{Fn}{Function}{:}{}
\SetKwFunction{FEvaluate}{Evaluate}
\SetKwFunction{Fapply}{apply}
\SetKwFunction{Funify}{unify}
\SetKwFunction{Fextend}{extend}
\SetKwIF{If}{ElseIf}{Else}{if}{then}{else if}{else}{endif}

\newcommand{\lift}[1]{{\uparrow}#1}

\SetCommentSty{mycommfont}
\usepackage{libertine}

\title{On The Suitability of Differential Dataflow For Datalog Interpretation In Highly Dynamic Settings}

\author{Bruno Rucy Carneiro Alves de Lima}
\orcid{0000-0002-2679-6639}
\affiliation{%
	\institution{University of Tartu}
	\department{Institute of Computer Science}
	\city{Tartu}
	\country{Estonia}
}
\email{bruno98@ut.ee}

\author{Merlin Kramer}
\orcid{0009-0000-1413-0680}
\affiliation{%
	\institution{Independent Researcher}
	\city{Wuppertal}
	\country{Germany}
}
\email{merlin.kramer@d10.pw}

\author{Kalmer Apinis}
\orcid{0009-0006-2395-6584}
\affiliation{%
	\institution{University of Tartu}
	\department{Institute of Computer Science}
	\city{Tartu}
	\country{Estonia}
}
\email{kalmera@ut.ee}

\begin{abstract}
	In the domain of knowledge representation and reasoning within AI, datalog engines play an ever-increasingly crucial role. The crux of their
	operation lies in materialization: the evaluation of a datalog program and its incorporation into a database. This operation becomes complex
	and resource-intensive, especially when the data is highly dynamic, as it is common in distributed environments. Thus, incremental materialization, adjusting
	the computation to new data instead of restarting it, is the norm. However, handling the deletion of data is significantly more complicated than addition
	due to the cascading effects of what is being removed. Differential Dataflow offers a computational model that effectively addresses this, ensuring
	consistent performance for both data additions and deletions. In this paper, we delve into the efficiency of materialization using three distinct
	datalog implementations: one based on a streamlined relational engine and two others that implement the same algorithm, but with one utilizing
	differential-dataflow, and another not. Our insights provide a roadmap for enhancing datalog-driven computations, particularly in dynamic data
	environments like the cloud.
\end{abstract}

\keywords{datalog, incremental view maintenance, differential dataflow}

\begin{document}

\maketitle

\section{Introduction}
Datalog\cite{datalog} is increasingly more relevant in the modern context. While it has traditionally been used almost exclusively for the workhorse of symbolic
AI, database reasoning, more recent implementations, such as the Rego\cite{rego} language for cloud policy evaluation, showcase the continued relevance and
flexibility of datalog.

The Open Policy Agent\cite{opa} (OPA) is a versatile policy engine designed to unify policy enforcement across the cloud-native stack. Central to OPA
is its policy language, Rego, which is a variant of datalog. Like Datalog, it is a declarative logic programming language, where instead of writing imperative
code, developers specify constraints, policies about cloud-components, and the engine figures whether these these policies have been violated or not, given a query.

The key use of Rego in OPA is to assert policies over structured data, most specifically over JSON. For instance, in the Kubernetes\cite{kubernetes} container
orchestrator, one might define a policy that prevents applications from being exposed externally unless certain annotations are present. This policy would be
represented as a set of Rego rules. When a request to the Kubernetes server occurs, the request would be evaluated against these rules, and the server would
allow or deny it based on the result.

Given the high velocity and volume of cloud-native data, it is paramount that policy decisions are made efficiently. This often implies adjusting an already-existing
evaluation, referred to as a materialization, to more facts, instead of scraping it and starting anew. While handling the addition of more facts is known to
be efficiently managed, deletions stand as an more complex challenge.

The naive approach to handling deletions, retracting a fact alongside everything else derived from it, might take longer than restarting the computation. This
inefficiency led to the emergence of the delete-rederive method\cite{dred}, that addresses this issue by computing the adjustment through the evaluation of
new datalog programs, first calculating all possible deletions, and then determining alternative derivations. The difference between these sets represents
the actual facts to be deleted.

Handling fact additions and retractions in different ways leads to drastically biased performance characteristics\cite{dredbf}. Because of this, modern datalog
engines tend to entirely avoid supporting the latter form of updates. Differential Dataflow\cite{differential_dataflow} is a general dataflow programming model
that lifts an iterative non-incremental algorithm into its incremental, referred to as differential, version. By outlining an algorithm over some input data with
differential dataflow, it will then work over update differences of the input data, that can be either positive, such as adding new data, or negative, as
removals. This could be leveraged in datalog evaluation as a way to handle updates in an uniform manner.

We note that incremental datalog evaluation is akin in process to online learning\cite{online_machine_learning}, and any results from the experiments presented
here are also indicative of the suitability of differential dataflow for problems in that sphere as well.

\textbf{Contributions.} We present an algorithm for interpreting datalog programs as a dataflow of facts and rules, with precise semantics. The semantic
medium for that algorithm is the DBSP language, that is able to formalise a limited subset of differential dataflow (DD). We then implement it in rust with
the differential dataflow library, and compare it with the most common methods of datalog evaluation, semi-naive evaluation, and the delete-rederive
method.

Our comparison provides a thorough overview on performance and memory usage over multiple programs, datasets and evaluation methods. We compared our algorithm
with one state of the art engine, and two off-the-shelf datalog engines, out of which one uses the method of converting a datalog program to relational algebra, and
another that uses a less common term rewriting method.

The state of the art engine that we chose to compare with is Soufflé\cite{souffle}. As most datalog engines, it does not support incremental evaluation, therefore
being a good example on how subpar their performance can be in highly dynamic scenarios.

\textbf{Structure of the paper.} The paper is organized as follows:
\begin{itemize}
	\item \textbf{Related Works} provides context to the paper by referencing recent high-profile works in the subfield of incremental datalog evaluation,
	      alongside other attempts at datalog evaluation that have used differential dataflow in some way.
	\item \textbf{Background} contains an overview of Datalog and its evaluation methods, including a detailed explanation of the semi-naive evaluation and how
	      it can be incrementally maintained.
	\item \textbf{Proposal} brings out our approach for modelling datalog interpretation with DBSP, how it relates to Differential Dataflow, and why that could
	      be beneficial for evaluating datalog programs in low-latency environments.
	\item \textbf{Experiments} details the empirical evaluation setup, including the datasets used, benchmarks, and a comparative analysis with existing systems. It
	      evaluates the performance of our approach in terms of runtime efficiency and memory usage.
	\item \textbf{Conclusion.} The final section concludes the paper by summarizing our findings, discussing the implications of our work, and suggesting
	      future directions for research in this area.
\end{itemize}

\section{Related Works}
The state-of-the-art of incremental evaluation of datalog programs is \cite{maintrevis}. It provides a thorough overview of the two most relevant classical methods, delete-rederive
and counting, alongside substantial empirical evaluations that provide key observations on their practical performance.

As the main novelty of this paper is in implementing a datalog interpreter with differential dataflow, we acknowledge the existence of two other datalog engines. The most
high profile one is DDLog\cite{ddlog}. It compiles a datalog program into a Rust DD program that executes recursive relational algebra, therefore being similar in strategy
to Soufflé. Our system is orthogonal to it, being an interpreter, hence not suffering from long compilation times, and in programs being freely changeable in runtime.

Laddder\cite{laddder} is an incremental evaluator of lattice-based datalog programs that extends differential dataflow. Our reasoner diverges in fundamental principles, as
it does not have lattice semantics, it is an interpreter, and focuses on evaluating general datalog programs.

\section{Background}
Datalog\cite{datalog} is a programming language with semantics denoting the evaluation of a set of possibly-recursive restricted horn clauses, a program, over
a fact store while remaining not turing complete. Evaluating a program $\Pi$ entails computing all implicit consequences over a fact store $E$, yielding new
facts. A Program is a set of rules of the following form: \[\textbf{h}(x_1, ..., x_j) \leftarrow \bigwedge_{i=1}^k\textbf{b}_i(x_1, ..., x_j)\] with $h$ as the
head atom, containing $x_i$ terms that can be either constant or variable, and $B = \bigwedge_{i=1}\text{b}_i$ as the body with each $b_i$ being an atom.

Materialization, the incorporation of all of a program's consequences $I = \Pi{(E)} \cup E$, eliminates the need for reasoning during query answering. Maintaining
this computation in face of additions $E^{+}$ and deletions $E^{-}$, $I' = \Pi{(I \cup E^{+} \// E^{-})} \cup (E \cup E^{+} \// E^{-})$ is the crux of modern Datalog, as
it relates to the broader problem of incremental view maintenance\cite{ivm} (IVM).

The standard procedure for obtaining $\Pi{(E)}$ is semi-naive evaluation\cite{datalog}. Evaluation is most frequently defined as the least fixed point (LFP) of the
immediate consequence $T{(E)}$\cite{datalog}. The immediate consequence of some program is the set of all facts which directly follow from the program over the data
i.e if $\Pi = \{\textbf{hop}(?x, ?y) \leftarrow \textbf{e}(?x, ?y), \textbf{e}(?y, ?z)\}$ and $E = \{\textbf{e}(a, b), \textbf{e}(b, c), \textbf{e}(c, d)\}$ then \[T = \{\textbf{hop}(a, c), \textbf{hop}(b, d)\}\]
which in this case $\Pi{(E)} \equiv T(E)$.
\subsection{Monotone evaluation}
Computing the LFP of $T$ can be done intuitively, by applying $T$ until there are no more new facts: $\Pi{(E)} = \bigcup_{i=1}^{n} T(I_{i})$, with $I_{i}$
representing the intermediate materialization. This method is called naive evaluation. Semi-naive evaluation differs in that the program is
transformed into a delta program $\Delta{\Pi}$, with new rules such that only the most recently inferred facts can progress the computation. Given a
$\Pi$ with rules $r_0, ..., r_n$ each having $k$ body atoms, up to $k$ new rules are created for each, with every $i$-th new rule containing
the same body atoms sans $b_i$, which is substituted for $\Delta{b_i}$, a set that only ever contains what has been inferred in the previous
recursion step.
\begin{example}{Semi-naive Evaluation Programs}
	\begin{align*}
		\Pi = \{         & \textbf{tc}(?x, ?y) \leftarrow \textbf{edge}(?x, ?y)                               \\
		                 & \textbf{tc}(?x, ?y) \leftarrow \textbf{tc}(?x, ?y), \textbf{tc}(?y, ?z)\}          \\
		\Delta{\Pi} = \{ & \textbf{tc}(?x, ?y) \leftarrow \textbf{edge}(?x, ?y)                               \\
		                 & \textbf{tc}(?x, ?z) \leftarrow \Delta{\textbf{tc}}(?x, ?y), \textbf{tc}(?y, ?z)    \\
		                 & \textbf{tc}(?x, ?z) \leftarrow \textbf{tc}(?x, ?y), \Delta{\textbf{tc}}(?y, ?z) \}
	\end{align*}
	\label{exsne}
\end{example}
Example \ref{exsne} contains the $\Delta{\Pi}$ of a transitive closure computation program. In $\Pi$, for the second rule, every iteration implies
seeking for matches between all of $\textbf{tc}$ and itself, hence implying an ever-increasing amount of redundant computations. With $\Delta{\Pi}$ however,
$\textbf{tc}$ is only joined with $\Delta{\textbf{tc}}$, minimizing redundancy with each iteration.
\subsection{Fact retractions}
Due to semi-naive evaluation's naturally incremental nature, it suffices for efficiently handling cases where $E^{-} = \emptyset$. The delete-rederive method\cite{dred} is
the standard for handling otherwise.
\begin{example}{The problem of overdeletion}
	\begin{align*}
		\Pi = \{   & \textbf{C}(?x) \leftarrow \textbf{A}(?x),   \\
		           & \textbf{C}(?x) \leftarrow \textbf{B}(?x) \} \\
		E =     \{ & A_0, \ldots, A_{n / 2}, B_0, \ldots, B_n \} \\
		E^{-} = \{ & A_0, \ldots, A_{n / 2} \}                   \\
		I =     \{ & C_{0}, \ldots, C_{n} \}                     \\
		I' =    \{ & C_{0}, \ldots, C_{n} \}
	\end{align*}
	\label{exod}
\end{example}
Example \ref{exod} demonstrates an instance of the problem of overdeletion, the first step of the two-step delete-rederive method. Here, program $\Pi$
comprises two rules that together infer $C$ either from the presence of facts in $A$ or $B$. Initially $E$ contains $n$ facts from $A$ and $B$, however
at some point, a deletion update $E^{-}$ arrives, containing all $A$ until $A_{n/2}$. DRED's expensive first step would unnecessarily delete all facts in
$C$ up until $C_n$, in spite of them still clearly holding due to $B$. This is partially addressed by the second part of the algorithm, that computes
alternative derivations of all overdeletion candidates.
\section{Proposal}
A promising way to compute updates to $\Pi(E)$ in a unified manner is to formalise it as a DBSP circuit, and incrementalize it. The main premise of DBSP is that any
algorithm described with it can be incrementalised with a deterministic algorithm. By doing so it is possible to take advantage of the efficient rust implementation
of DD\cite{rust_ddflow}, hence implementing naive datalog evaluation would yield semi-naive evaluation that could support retractions too.

\subsection{Substitution-based immediate consequence}
The most impactful aspect of evaluation is the implementation of $T(E)$. There are two main methods to do so. The first is to evaluate rules as a term rewriting
problem, named as the substitution-based method, and the second is to rewrite a datalog rules as relational algebra equations, and delegate their execution to an
efficient relational engine. The latter is the most popular, as it offloads significant amounts of complexity away. We focus however on the former, since it has had
less recent research interest, and is significantly easier to implement and parallelize.

The basis of the substitution-based method are substitutions $\sigma$\[\sigma = \{x_1 \mapsto y_1, ..., x_i \mapsto y_i\}\] with all $x$ being variable terms and all
$y$ constant. There are three fundamental operations, with $a$ and $b$ as atoms, and $\sigma_i$ as substitutions:
\begin{itemize}
	\item \textbf{Application}. apply($\sigma, a$):

	      The application of $\sigma$ to $a$ results in replacing every variable term $t$ in $a$ with its corresponding substitution in $\sigma$, if it
	      exists: \[\begin{cases} y_i & \text{if } t == x_i \\ t & \text{otherwise} \end{cases}\]Let $P$ be the symbol of $a$: \[\text{apply}(\sigma, a) = \textbf{P}(y_1, ..., y_n)\]
	      The result of application is then an atom with possibly less variables.
	\item \textbf{Composition}. ($\sigma_1 \circ \sigma_2$): $\text{apply}(\sigma_1, \text{apply}(\sigma_2, x))$
	\item \textbf{Extension}. extend($\sigma_1, \sigma_2$):

	      Extending $\sigma_1$ with $\sigma_2$ implies combining the mappings of both substitutions:\[\text{extend}(\sigma_1, \sigma_2) = \begin{cases} \sigma_1 \cup \sigma_2 & \text{if } \forall x, \sigma_1(x) = \sigma_2(x) \\ \text{otherwise} \end{cases}\]. The first case is a simple union of the two substitutions because they're consistent (they don't provide different mappings for the same variable), and
	      the second case handles that if there's any variable for which the two substitutions provide different mappings, the extend operation fails.
	\item \textbf{Unification}. unify($a, b$):

	      Unification is an attempt to find a $\sigma$ such that when it is applied to some atom $a$ and another atom $b$ that only has constants, the resulting atoms are identical.
	      \[
		      \text{unify}(a, b) =
		      \begin{cases}
			      \sigma      & \text{if } \exists \sigma \text{ such that } \sigma(a) = \sigma(b) \\
			      \text{fail} & \text{otherwise}
		      \end{cases}
	      \]
\end{itemize}
The substitution-based method relies on computing the union of the immediate consequence of each rule. This is done as follows:

Define the initial set of substitutions as $\Sigma_0 = \{ \sigma_0 \}$, where $\sigma_0$ is an empty substitution. For
each body atom $b_m$, find the set of ground facts $F_j \subseteq F$ that match $b_m$.
\begin{algorithm}[h]
	\small
	\DontPrintSemicolon
	\SetNoFillComment
	\Input{A body atom list \(B\), set of ground facts \(E\), head atom \(h\)}
	\Output{Immediate Consequence \(I\)}

	\Fn{\FEvaluate{\(b_i, E, \Sigma_{i-1}\)}}{
		\(A = \{\}\) \tcc*{Initialize the set of fresh atoms}

		\ForEach{\(\sigma \in \Sigma_{i-1}\)}{
			\(A \cup (\sigma, \Fapply{$\sigma, b_i$})\)
		}

		\(\Sigma_i = \emptyset\) \tcc*{Initialize the set of expanded substitutions}

		\ForEach{\((\sigma, a) \in A\)}{
			\ForEach{fact \(f \in E\)} {
				\(\sigma' \gets \Funify{$a, f$}\) \tcc*{Unify the current fresh atom with the fact}
				\If{\(\sigma' \neq \text{fail}\)}{
					\(\sigma'' \gets \Fextend{$\sigma, \sigma'$}\) \tcc*{Extend its respective substitution}
					\If{\(\sigma'' \neq \text{fail}\)}{
						\(\Sigma_i \gets \Sigma_i \cup \{\sigma''\}\) \tcc*{Add to the current set}
					}
				}
			}
		}

		\If{\(i+1 \leq |B|\)}{
			\Return \(\FEvaluate{$b_{i+1}, E, \Sigma_i$}\) \tcc*{Process the next body atom}
		}
		\Else{
			\Return \(\Sigma_m\) \tcc*{Return the final set of substitutions}
		}
	}

	\(I \gets \emptyset\) \tcc*{Initialize the immediate consequence set}
	\(\Sigma_0 \gets \{ \text{empty substitution} \}\) \tcc*{Initial set of substitutions}

	\(\Sigma_m \gets \FEvaluate{$b_1, E, \Sigma_0$}\) \tcc*{Start with the first body atom}

	\For{each final substitution \(\sigma_m \in \Sigma_m\)}{
		\(I \gets I \cup \{\Fapply{$\sigma_m, h$}\}\) \tcc*{Apply the substitutions}
	}

	\caption{Substitution-based Immediate Consequence}
	\label{alg:recursive_substitution}
\end{algorithm}

Algorithm \ref{alg:recursive_substitution} is the standard substitution-based method. We exemplify its execution on table \ref{tab:substitution_iterations}.

\begin{table}
	\small
	\caption{Substitution-based method iterations}
	\centering
	\begin{tabular}{|c|c|c|c|c|}
		\hline
		\textbf{i}         & \textbf{Body Atom}           & \textbf{Substitution}                          & \textbf{Fresh Atom} & \textbf{Unif} \\
		\hline
		\multirow{2}{*}{1} & \multirow{2}{*}{$R(?x, ?z)$} & $\{\}$                                         & $R(a, c)$           & success       \\
		                   &                              & ...                                            & $R(b, d)$           & success       \\
		\hline
		\multirow{6}{*}{2} & \multirow{6}{*}{$T(?x, ?y)$} & $\{?x \mapsto a, ?z \mapsto c\}$               & $T(a, b)$           & success       \\
		                   &                              & ...                                            & $T(b, c)$           & fails         \\
		                   &                              & ...                                            & $T(c, d)$           & fails         \\
		                   &                              & $\{?x \mapsto b, ?z \mapsto d\}$               & ...                 & fails         \\
		                   &                              & ...                                            & ...                 & success       \\
		                   &                              & ...                                            & ...                 & fails         \\

		\hline
		\multirow{6}{*}{3} & \multirow{6}{*}{$T(?y, ?z)$} & $\{?x \mapsto a, ?y \mapsto b, ?z \mapsto c\}$ & $T(a, b)$           & fails         \\
		                   &                              & ...                                            & $T(b, c)$           & success       \\
		                   &                              & ...                                            & $T(c, d)$           & fails         \\
		                   &                              & $\{?x \mapsto b, ?y \mapsto c, ?z \mapsto d\}$ & ...                 & fails         \\
		                   &                              & ...                                            & ...                 & fails         \\
		                   &                              & ...                                            & ...                 & success       \\
		\hline
	\end{tabular}
	\label{tab:substitution_iterations}
\end{table}

The non-trivial program is: \[\Pi = \{\textbf{S}(?x, ?z) \leftarrow \textbf{R}(?x, ?z), \textbf{T}(?x, ?y), \textbf{T}(?y, ?z)\}\] with fact store $E = \{\textbf{T}(a, b), \textbf{T}(b, c), \textbf{T}(c, d), \textbf{R}(a, c), \textbf{R}(b, d)\}$. The
final substitutions(those that will be applied to the head of their respective rules) are \[\{?x \rightarrow a, ?y \rightarrow b, ?z \rightarrow c\}, \{?x \rightarrow b, ?y \rightarrow c, ?z \rightarrow d\}\] therefore
inferring two atoms: $\textbf{S}(a, c)$ and $\textbf{S}(b, d)$. Column \textbf{Unif} shows all unification and substitution extension extension attempts. As it can be
seen, there is a large amount of wasteful attempts. This can however, be partially remedied by extensive parallelism, since every attempt in the same iteration $i$ can
happen independent of each other. Aside from incrementalisation, DD's implementation offers extensive facilities for computation distribution, leveraging from its
underlying model, Timely Dataflow\cite{timely}.
\subsection{The issue}
We can then summarise Algorithm \ref{alg:recursive_substitution}'s problems:
\begin{enumerate}
	\item \textbf{Complexity}: The nested loops combined with recursive calls can lead to high computational overheads. For each body atom, the algorithm attempts to unify against every fact
	      in the database and every existing partial substitution. As the number of facts or the length of the rule body grows, this can become extremely costly.
	\item \textbf{Memory Usage}: $\Sigma_i$, can grow rapidly, especially if there are many possible substitutions that satisfy the body atoms.
	\item \textbf{Redundancy}: Many of the computed substitutions might be redundant or irrelevant for deriving the final consequence, leading to unnecessary computational work.
	\item \textbf{Lack of Indexing}: The method as presented does not leverage any indexing on the facts. In practice, efficient indexing techniques can vastly speed up the lookup operations, making
	      rule evaluations much faster.
	\item \textbf{Lack of Optimization}: Traditional relational database systems employ various optimization techniques, like query rewriting, join ordering, and cost-based optimization. The pure
	      substitution-based method does not inherently take advantage of these techniques, since it is not built on top of a relational engine.
	\item \textbf{Scalability Issues}: Techniques like data partitioning and distributed computation, which are often used in modern Datalog systems, aren't straightforwardly integrated into this model.
\end{enumerate}

Relational engines do not suffer from from almost every single mentioned limitation. By utilizing DD however, we can retain the simplicity of substitution-based evaluation while overcoming a significant
portion of the downsides, and obtaining the following advantages:
\begin{itemize}
	\item \textbf{Incremental Computation}: DD processes changes to data rather than recomputing results from scratch. This is particularly beneficial in the context of Datalog evaluation, where small
	      updates to the facts or rules might necessitate large recalculations.
	\item \textbf{Optimized Memory Usage}: Instead of storing entire sets of substitutions, DD could maintain differences between sets of, which can lead to significant memory savings.
	\item \textbf{Efficient Indexing}: Arrangements are partitioned indices, implemented as in-memory LSM trees, designed for efficient data organization, enabling rapid lookups, joins, and aggregation. These
	      indices facilitate the persistent reuse of indexed views.
\end{itemize}
\subsection{A DBSP circuit for substitution-based evaluation}
DBSP is a language that provides a unified, formal and sound foundation for manipulating non-streaming, streaming and incremental computations. It has been shown\cite{dbsp} to model many rich query
languages, such as relational algebra extended with aggregates and both monotonic and non-monotonic recursion. The main assumption is that inputs must arrive in time order, which allows for nested
time domains. This restricts it to a subset of DD.

Streams $S$ are maps $\mathbb{N} \rightarrow A$ with the domain representing time, and the codomain a value from an abelian group $A$, within the context of this work. Stream operators are
functions $S_A \rightarrow S_B$, extending to multiple inputs and outputs. The most elementary operator is stream lifting $\lift$. Given a function $f : A \rightarrow B$, $\lift{f} : S_A \rightarrow S_B$.

The delay operator $z^{-1}_A: S_A \rightarrow S_A$ produces a stream that is one step behind its domain, and is fundamental to implementing both naive and semi-naive evaluation. Stream operators
are time invariant, if $S \circ z^{-1} \equiv z^{-1} \circ S$. The most notable property is causality. An operator is \textit{causal} if it's output at time $t$ only ever relies on all inputs with
time $t'$ such that $t' \leq t$, with it being \textit{strict} if $t' < t$.

Lifted operators are not necessarily strict, but are causal. The main requirement for implementing algorithm \ref{alg:recursive_substitution} is in computing the least fixed point of the immediate
consequence operator.

Given that all strict operators are guaranteed to have a unique solution to their fixpoint, and assuming that the immediate consequence could be outlined as a causal operator, then
due to the fact that for every $F : S_B \rightarrow S_B$ that is strict, and $T : S_A \times S_B \rightarrow $ that is causal, it follows from $\textbf{fix } \alpha.(T \circ F) \,\alpha$ being well-defined
and causal, that if algorithm \ref{alg:recursive_substitution} is $T$, and the delay operator is $F$, then the least fixed point of the immediate consequence is well-defined.

Linear operators are those that are group homomorphisms, with Bi-linear ones extending that concept to operators with multiple inputs. All linear operators are causal. We compose algorithm \ref{alg:recursive_substitution}'s circuit entirely with causal operators and functions, therefore it being causal itself. Furthermore, as referenced in \cite{dbsp}, we model facts, rules
and substitutions as the $\mathbb{Z}$-set abelian group, which allows for the usage of relational operators with correct set semantics, by utilizing the $\textit{distinct}$ operator. $\textit{distinct}$'s $\mathbb{Z}$-set semantics coalesces values with negative multiplicities into set removals, and ensures that every value has multiplicity of at most one.

\begin{figure}[!htbp]
	\centering
	\caption{Datalog interpretation as a DBSP circuit \label{fig:dbsp-circuit}}
	\begin{tikzpicture}[node distance=1.5cm,>=latex]
		\node[] (subs) {$\Sigma_0$};
		\node[right of=subs, shape=circle, inner sep=0pt, node distance=1cm] (subcat) {$+$};
		\node[block, below of=subcat, node distance=0.75cm] (sr) {$\bowtie_{(rule\_id,pos)}$};
		\node[below of=subs] (rules) {$\Pi$};
		\node[block, below of=sr, node distance=2.0cm] (ids) {$\lift{\textit{identifiers}}$};
		\node[block, below of=sr, node distance=1.0cm] (m1) {$m_1$};
		\node[block, right of=m1, node distance=1.25cm] (m3) {$\lift{m_3}$};
		\node[block, right of=m3, node distance=1.25cm] (fs) {$\bowtie_{sym}$};
		\node[block, above of=fs, node distance=1.0cm] (m4) {$m_4$};
		\node[block, above of=m3, node distance=1.75cm] (m4delay) {$z^{-1}$};
		\node[block, right of=fs, node distance=1.75cm] (sh) {$\bowtie_{(rule\_id,pos)}$};
		\node[block, right of=ids, node distance=2.0cm] (m2) {$m_2$};
		\node[block, right of=sh, node distance=1.5cm] (m5) {$m_5$};

		\node[below of=rules, node distance =2.0cm] (facts) {$E$};
		\node[below of=m5, shape=circle, inner sep=0pt, node distance =0.75cm] (factcat) {$+$};
		\node[block, right of=factcat, node distance=1cm] (finaldist) {$\lift{\textit{distinct}}$};
		\node[block, below of=finaldist, node distance=1.20cm] (factdelay) {$z^{-1}$};
		\node[right of=m4, node distance =1.0cm, inner sep=0pt] (ghost) {};

		\node[above right of=factdelay, node distance =0.90cm] (out) {$I$};

		\draw[->] (subs) -- (subcat);
		\draw[->] (m3) -- (fs);
		\draw[->] (fs) -- (m4);
		\draw[->] (m4) -| (sh);
		\draw[->] (ghost) |- (m4delay);
		\draw[->] (m4delay) -- (subcat);
		\draw[->] (sh) -- (m5);
		\draw[->] (m5) -- (factcat);
		\draw[->] (subcat) -- (sr);
		\draw[->] (facts) -| (factcat);
		\draw[->] (factcat) -- (finaldist);
		\draw[->] (finaldist) -- (factdelay);
		\draw[->] (factdelay) -| (fs);
		\draw[->] (rules) |- (ids);
		\draw[->] (ids) -- (m1);
		\draw[->] (m1) -- (sr);
		\draw[->] (ids) -- (m2);
		\draw[->] (m2) -| (sh);
		\draw[->] (sr) -| (m3);
		\draw[->] (finaldist) |- (out);
	\end{tikzpicture}
\end{figure}
Figure \ref{fig:dbsp-circuit} showcases the DBSP circuit of algorithm \ref{alg:recursive_substitution}. It implements naive evaluation, therefore not being incremental. DBSP however stands
out for its unique capability to deterministically incrementalize any circuit. When this circuit is outlined with the DD library, it undergoes automatic incrementalization and parallelization, enhancing
its efficiency by handling differences of facts, rules, and substitutions.
\begin{figure}[!htbp]
	\centering
	\caption{Figure \ref{fig:dbsp-circuit}'s operator semantics \label{fig:dbsp-circuit-semantics}}
	\begin{align*}
		\textbf{m}_0 & : \lambda (rule\_id,rule).                                              \\
		             & \quad \textbf{map}((\lambda(pos, atom).                                 \\ & \quad \quad (((rule\_id,pos),atom), 1)), \textbf{enumerate}(\textbf{snd}(rule)))                                                                     \\
		\textbf{m}_1 & : \lambda rules.(\textbf{flatmap}(m_0, rules))                          \\
		\textbf{m}_2 & : \lambda rules.                                                        \\ & \quad (\lambda (rule\_id, rule).((rule\_id, \textbf{len}(\textbf{snd}(rule))), \textbf{fst}(rule)) rules)                                                             \\
		\textbf{m}_3 & : \lambda (heads, subs).                                                \\ & \quad (map (\lambda ((key, head), (\_key, sub)).\\ & \quad \quad (\textbf{fst}(head), (key, (sub, \textbf{apply}(sub, head)))) ))                                            \\
		\textbf{u}   & : \lambda (\text{rewrite, fact}). \textbf{unify}(\text{rewrite, fact})  \\
		\textbf{e}   & : \lambda (\text{sub, new\_sub}). \textbf{extend}(\text{sub, new\_sub}) \\
		\textbf{o}_1 & : \lambda(fact,(sub,rewrite)).                                          \\ & \quad \textbf{flatmap}((\lambda new\_sub.\textbf{e}(sub,new\_sub)),\textbf{u}(rewrite,fact))                                                            \\
		\textbf{o}_2 & : \lambda(fact,(rule\_id,pos),(sub,rewrite)).                           \\ & \quad \textbf{flatmap}(\lambda e\_sub.\\ & \quad \quad  ((rule\_id,(+\,pos\,1)),e\_sub)),\textbf{o}_1(fact, (sub, rewrite))                                \\
		\textbf{m}_4 & : \lambda(facts, headsXsubs).                                           \\ & \quad \textbf{map}((\lambda(fact,(\_symbol,((rule\_id,pos),(sub,rewrite)))).\\ & \quad \quad \textbf{o}_2(fact,rule\_id,pos,sub,rewrite)),\textbf{zip}(facts,headsXsubs)) \\
		\textbf{m}_5 & : \lambda(heads, new\_subs).                                            \\ & \quad (\lambda (((\_rule\_id, \_rule\_len), head), ((\_rule\_id, pos), sub) ).\\ & \quad \quad \textbf{apply}(sub, head))
	\end{align*}
\end{figure}

In figure \ref{fig:dbsp-circuit-semantics}, we precisely specify the complex semantics of operators $m_1, ..., m_5$ in lambda calculus, and then in natural language. These
operators are integral components of the circuit, with their unlifted versions translating the imperative style of algorithm \ref{alg:recursive_substitution} into a
functional one, more suitable to dataflow computation. We employ extensive usage of standard scalar functions from the functional programming literature, readily available
in the DD library. Operator $\textbf{m}_0$ processes a rule identifier and a rule, enumerating the atoms in the rule and associating each with its position and the rule
identifier, each combination having a multiplicity of 1, signalling that iteration for new substitutions starts at the first body atom, and procedes atom-by-atom. $\textbf{m}_1$
applies it to all incoming rules by flatmap-ing them with $\textbf{m}_0$.

$\textbf{m}_2$ pairs each rule's identifier with a tuple containing the head of the rule and the length of its body, for every rule in a given $\mathbb{Z}\text{-set}$ of
rules. $\textbf{m}_3$ is tasked with taking pairs of heads and substitutions and mapping over them to produce a new fact, as the result of applying a substitution of a rule head.

The $\textbf{u}$ operator is responsible for unifying rewrites with facts, while $\textbf{e}$ extends a substitution with a new one. $\textbf{o}_1$ and $\textbf{o}_2$ are
more complex, involving the unification of atoms with facts and the extension of original substitutions, with $\textbf{o}_2$ also incrementing the position and associating
it with the rule identifier, therefore either invalidating substitutions, or forwarding them to the next iteration.

$\textbf{m}_4$ maps over pairs of old facts and heads/substitutions tuples, applying $\textbf{o}_2$ to produce a sequence of extended substitutions. $\textbf{m}_5$ works
with pairs of heads and new substitutions, applying each substitution to the respective head, getting ground facts.

With the circuit and its constituents outlined and described, we benchmark its implementation against multiple datalog engines.
\section{Experiments}
Two experiments were conducted in order to compare incremental reasoning performance. The first, measured materialization adjustment latency under both additions and deletions. The
second recorded the maximum memory resident set size under full materialization. We refer to the reasoner implemented with DD as \textbf{Diff}. We also compare with other reasoners,
out of which two were implemented by us, due to a lack of open-source incremental reasoners. We name the first \textbf{Chibi}, and the second \textbf{Relational}. The first uses
the same substitution-based method as \textbf{Diff}, but not DD. The second, as its name hints at, is built on top of a relational engine implemented from scratch. Both are written
in Rust, and share as much code as reasonable between themselves and \textbf{Diff}, for the sake of fairness.

We also compare against a state of the art reasoner, Souffle. It can be run both in interpreted and compiled modes. We run it as an interpreter, as all of our reasoners are interpreters
too. Soufflé does not support incremental updates. Benchmarking it in how it could be used in a dynamic setting is very valuable however, given that most reasoners are not incremental,
therefore being exemplary as to how inefficient those are in dynamic settings, which are more realistic than static scenarios.

\textbf{Setup.} Experiments were run on a amazon-web-services-provisioned m1 mac machine, with 8 cpu cores and 16 gigabytes of RAM. DDLog was version 1.2.3, Souffle 2.4,
and Rust 1.72. Each benchmark measurement was taken a sufficient number of times to ensure that variance was low. No other user-space processes aside from the benchmark
was running.
\begin{table}[!htbp]
	\centering
	\caption{Dataset Overview}
	\begin{tabular}{|c|c|c|}
		\hline
		Dataset & Area of Interest & Programs      \\
		\hline
		LUBM    & semantic web     & RhoDF, OWL2RL \\
		\hline
		RMAT1K  & synthetic        & tc            \\
		\hline
		RAND1K  & synthetic        & tc            \\
		\hline
	\end{tabular}
	\label{tab:datasets}
\end{table}

\textbf{Datasets.} On table \ref{tab:datasets} all datasets and program names are shown. There are two areas of interest. The semantic web has pushed
the datalog envelope by being an extensive source of improvements to already-established algorithms\cite{dredbf}, and in providing a myriad of both
synthetic and real datasets. The second area of interest is of graph benchmarks that follow some distribution, allowing for the measurement of one
of the most important aspects of a recursive datalog evaluation, the cost of iteration.
\begin{itemize}
	\item \textbf{LUBM} is a known benchmark dataset for both RhoDF\cite{rhodf} and OWL2RL rulesets. The data is divided in two parts, the ontology that describes
	      universities, and the assertional data, that contains facts about universities describes with the ontology. RhoDF is a single-relation program that realizes graph
	      entailment of a subset of the RDFS\cite{rdfs} metamodel. All rules are mutually recursive and have the same head, hence being inefficient to evaluate and
	      parallelise. OWL2RL\cite{owl2rl} is a description logic, a subset of the OWL language, that is meant to be implemented with rules-based languages such as
	      datalog. This program was generated by converting OWL2RL entailments to datalog rules\cite{descr_to_datalog}. It has over 100 rules, being a canary for reasoner efficiency.
	\item \textbf{RMAT10k} is a standard benchmark used by various other reasoners\cite{recstep,bigdatalog,nexus,cog}. It is a dense graph that follows the \verb|rmat| profile of the
	      GT\cite{gtgraph} synthetic graph benchmark. It contains 10x as many edges as vertices, that follow a power-law distribution.
	\item \textbf{RAND1k} is a graph from GT as well, with 1000 nodes that each have a 1\% chance of being connected to every other.
\end{itemize}

\subsection{Runtime comparison}
Tables \ref{tab:mainbenchmark1} and \ref{tab:mainbenchmark2} show the results of the first benchmark. Three measurements were taken, that each recorded the time to materialize, respectively, $E$, the
initial materialization batch, $E^{+}$, the adjustment to an update with additions, and $E^{-}$, adjusting to deleting the same data that was added with $E^{+}$. All measurements
are in \textbf{seconds}.
\begin{table}[!htbp]
	\centering
	\caption{Runtime Experimental Results - I}
	\scriptsize
	\begin{tabular}{|c|c|c|c|c|c|c|c|c|c|c|c|c|c|c|c|c|c|}
		\hline
		\multirow{2}{*}{\textbf{Dataset}} & \multirow{2}{*}{\textbf{Program}} & \multirow{2}{*}{\textbf{Batch}} & \multicolumn{3}{c|}{$\textbf{Diff}$} & \multicolumn{3}{c|}{\textbf{Chibi}}                                    \\
		\cline{4-9}
		                                  &                                   &                                 & $E$                                  & $E^+$                               & $E^{-}$ & $E$  & $E^+$ & $E^{-}$ \\
		\hline
		\multirow{10}{*}{LUBM1}           & \multirow{6}{*}{rdfs}             & 50\%                            & 0.27                                 & 0.64                                & 0.65    & 0.30 & 0.50  & 0.40    \\
		                                  &                                   & 75\%                            & 0.41                                 & 0.48                                & 0.50    & 0.45 & 0.46  & 0.32    \\
		                                  &                                   & 90\%                            & 0.50                                 & 0.40                                & 0.39    & 0.52 & 0.45  & 0.30    \\
		                                  &                                   & 99\%                            & 0.53                                 & 0.33                                & 0.32    & 0.57 & 0.46  & 0.28    \\
		                                  &                                   & 100\%                           & 0.89                                 & 0                                   & 0       & 0.57 & 0     & 0       \\
		\cline{2-9}
		                                  & \multirow{5}{*}{owl2rl}           & 50\%                            & 1.62                                 & 4.43                                & 4.60    & 6.40 & OOT   & OOT     \\
		                                  &                                   & 75\%                            & 3.33                                 & 2.62                                & 2.68    & 15.3 & OOT   & OOT     \\
		                                  &                                   & 90\%                            & 5.00                                 & 1.04                                & 1.01    & 22.4 & OOT   & OOT     \\
		                                  &                                   & 99\%                            & 5.83                                 & 0.01                                & 0.01    & 29.8 & OOT   & OOT     \\
		                                  &                                   & 100\%                           & 5.82                                 & 0                                   & 0       & 31.1 & 0     & 0       \\
		\hline
		\multirow{5}{*}{RAND-1k}          & \multirow{5}{*}{tc}               & 50\%                            & 0.02                                 & 0.49                                & 0.22    & 0.02 & 0.36  & 0.72    \\
		                                  &                                   & 75\%                            & 0.09                                 & 0.42                                & 0.42    & 0.10 & 0.30  & 1.53    \\
		                                  &                                   & 90\%                            & 0.28                                 & 0.25                                & 0.24    & 0.33 & 0.35  & 4.51    \\
		                                  &                                   & 99\%                            & 0.48                                 & 0.06                                & 0.05    & 0.57 & 0.37  & 0.52    \\
		                                  &                                   & 100\%                           & 0.53                                 & 0                                   & 0       & 0.63 & 0     & 0       \\
		\hline
		\multirow{5}{*}{RMAT-1k}          & \multirow{5}{*}{tc}               & 50\%                            & 37.0                                 & 138                                 & 95.1    & 69.4 & 46.7  & 104     \\
		                                  &                                   & 75\%                            & 68.2                                 & 79.2                                & 67.2    & 83.1 & 24.8  & 73.1    \\
		                                  &                                   & 90\%                            & 83.1                                 & 36.3                                & 35.1    & 92.8 & 10.7  & 47.5    \\
		                                  &                                   & 99\%                            & 100                                  & 5.11                                & 4.87    & 99.3 & 10.8  & 50.8    \\
		                                  &                                   & 100\%                           & 103                                  & 0                                   & 0       & 103  & 0     & 0       \\
		\hline
	\end{tabular}
	\label{tab:mainbenchmark1}
\end{table}
As an example, If the initial materialization batch size is 75\%, then $E$ represents the amount of time taken to materialize 75\% of the data, $E^{+}$ is how much the incremental materialization
of the remaining 25\% of the data took, and lastly, $E^{-}$ is how long did it take to adjust the materialization to the removal of that same update, that is, back to the same state that it was with $E$.

\begin{table}[!htbp]
	\centering
	\caption{Runtime Experimental Results - II}
	\scriptsize
	\begin{tabular}{|c|c|c|c|c|c|c|c|c|c|c|c|c|c|c|c|}
		\hline
		\multirow{2}{*}{\textbf{Dataset}} & \multirow{2}{*}{\textbf{Program}} & \multirow{2}{*}{\textbf{Batch}} & \multicolumn{3}{c|}{\textbf{Relational}} & \multicolumn{3}{c|}{\textbf{Souffle}}                                    \\
		\cline{4-9}
		                                  &                                   &                                 & $E$                                      & $E^+$                                 & $E^{-}$ & $E$  & $E^+$ & $E^{-}$ \\
		\hline
		\multirow{10}{*}{LUBM1}           & \multirow{6}{*}{rdfs}             & 50\%                            & 0.17                                     & 0.27                                  & 0.32    & 0.13 & 0.25  & 0.13    \\
		                                  &                                   & 75\%                            & 0.26                                     & 0.27                                  & 0.31    & 0.19 & 0.25  & 0.19    \\
		                                  &                                   & 90\%                            & 0.32                                     & 0.29                                  & 0.32    & 0.22 & 0.25  & 0.22    \\
		                                  &                                   & 99\%                            & 0.35                                     & 0.29                                  & 0.31    & 0.24 & 0.25  & 0.24    \\
		                                  &                                   & 100\%                           & 0.29                                     & 0                                     & 0       & 0.25 & 0     & 0       \\
		\cline{2-9}
		                                  & \multirow{5}{*}{owl2rl}           & 50\%                            & 11.8                                     & 40.1                                  & 6.01    & 0.12 & 0.16  & 0.12    \\
		                                  &                                   & 75\%                            & 38.3                                     & 36.6                                  & 5.98    & 0.13 & 0.16  & 0.13    \\
		                                  &                                   & 90\%                            & 58.1                                     & 31.0                                  & 5.93    & 0.15 & 0.16  & 0.15    \\
		                                  &                                   & 99\%                            & 70.9                                     & 12.7                                  & 0.20    & 0.16 & 0.16  & 0.16    \\
		                                  &                                   & 100\%                           & 73.7                                     & 0                                     & 0       & 0.16 & 0     & 0       \\
		\hline
		\multirow{5}{*}{RAND-1k}          & \multirow{5}{*}{tc}               & 50\%                            & 0.01                                     & 0.10                                  & 0.07    & 0.01 & 0.10  & 0.01    \\
		                                  &                                   & 75\%                            & 0.02                                     & 0.08                                  & 0.10    & 0.03 & 0.10  & 0.03    \\
		                                  &                                   & 90\%                            & 0.06                                     & 0.08                                  & 0.12    & 0.07 & 0.10  & 0.07    \\
		                                  &                                   & 99\%                            & 0.09                                     & 0.09                                  & 0.10    & 0.09 & 0.10  & 0.09    \\
		                                  &                                   & 100\%                           & 0.10                                     & 0                                     & 0       & 0.10 & 0     & 0       \\
		\hline
		\multirow{5}{*}{RMAT-1k}          & \multirow{5}{*}{tc}               & 50\%                            & 0.93                                     & 1.41                                  & 2.58    & 0.19 & 0.41  & 0.19    \\
		                                  &                                   & 75\%                            & 1.33                                     & 1.92                                  & 3.02    & 0.30 & 0.41  & 0.30    \\
		                                  &                                   & 90\%                            & 1.73                                     & 2.30                                  & 3.69    & 0.35 & 0.41  & 0.35    \\
		                                  &                                   & 99\%                            & 2.03                                     & 2.75                                  & 4.03    & 0.40 & 0.41  & 0.40    \\
		                                  &                                   & 100\%                           & 2.04                                     & 0                                     & 0       & 0.41 & 0     & 0       \\
		\hline
	\end{tabular}
	\label{tab:mainbenchmark2}
\end{table}
The choice of facts can significantly impact reasoning performance. However, conducting comprehensive performance assessments over all subsets is not feasible. This
comes from the extensive time required to execute the complete benchmark and the overwhelming factorial quantity of potential data permutations. To address this, we opted
for a streamlined approach. We randomly selected subsets, specifically choosing sizes that constitute 50\%, 25\%, 10\%, and 1\% of the original data. We note that this was
done differently for LUBM, since it has two components, TBox and ABox. We ensured that the TBox was always fed (fully materialized) first.

Since Soufflé does not support incremental additions nor deletions, it must rematerialize with each update. In order to compute the time taken to adjust to an update $E^{+}$, it is
necessary to compute the full materialisation of $E \cup E^{+}$. Conversely, to do so in place of $E^{-}$, would be the same as measuring $E \cup E^{+} \// E^{-}$, which is just $E$.

The first benchmark group consists of the two entailment programs that run on the LUBM dataset. In spite of RDFS having six rules and OWL2RL almost 100, the size of output $I$ is
similar. For RhoDF, \textbf{Diff} exhibits the expected uniformity in the handling of updates. The time taken for $E^{+}$ and $E^{-}$ are almost the same, with scalability seeming to
be linear to the impact of the size of the update. Soufflé materializes fast, and is able to be faster than all reasoners, including those that use DRED. Rematerialization however
is up to 25x slower than incremental addition for OWL2RL. We posit that this happens for \textbf{Diff} because OWL2RL is friendlier to parallelization than RhoDF, with it being
guided by the number of relations, out of which RhoDF has none, and OWL2RL has many.

We contrast this to \textbf{Relational} and \textbf{Chibi}, that each have parallelization on the level of rules, with each rule being applied in parallel. These two reasoners
furthermore, perform in opposite directions on OWL2RL, with \textbf{Chibi} running out of time (OOT) to compute additions and deletions, and \textbf{Relational} demonstrating
uniform behavior only with very small update sizes.

The second benchmark group focuses on measuring performance over a simple reachability program on graphs that are highly connected,x offering an overview on the effectiveness
of every reasoner's handling of long iterative chains. RAND-1k yields a small output, with only approximately 12000 facts being emitted. It does not pose a challenge to any
reasoner. RMAT-1k on the other hand, in spite of being ten times bigger, producs over 25x more facts with certain simple paths being thousands of edges long. This causes a
large amount of timestamp flux in differential dataflow, due to too-fine-grained difference tracking. Soufflé shines in this situation, with rematerialization reigning over
all update sizes aside from 1\% of less.
\subsection{Peak memory usage comparison.}
In this subsection we compare the maximum resident memory set during full materialization.
\begin{table}[!htbp]
	\caption{Memory usage experimental results}
	\begin{center}
		\begin{tabular}{|c|c|c|c|c|c|c|}
			\hline
			Dataset                & \textbf{Program} & \textbf{Diff} & \textbf{Chibi} & \textbf{Relational} & \textbf{Souffle} \\
			\hline
			\multirow{3}{*}{LUBM1} & rhoDF            & 488           & 576            & 421                 & 30               \\
			                       & owl2rl           & 446           & 6149           & 6840                & 20               \\
			\cline{2-4}
			\hline
			RAND-1k                & tc               & 90            & 41             & 13                  & 20               \\
			\hline
			RMAT-1k                & tc               & 5492          & 1409           & 1362                & 20               \\
			\hline
		\end{tabular}
	\end{center}
	\label{tab:memorybenchmark}
\end{table}
Table \ref{tab:memorybenchmark} presents the maximum resident memory set for each of the methods and programs across different datasets. Memory usage is presented in Mbs. LUBM1
takes up 20 Mbs of disk space, RAND-1k and RMAT-1k, respectively, 0.10Mbs and 1 Mb.

Soufflé consistently shows the lowest memory usage, highlighting the cost of incrementalization. It is at two times more memory efficient than all others, up to
25 times in RMAT. \textbf{Diff}'s poor handling of highly incremental scenarios is furthered by RMAT's memory consumption. \textbf{Diff} consumes 5000x more memory
than the input size. We profiled that the difference set that stores substitutions starts to grow quadratically as the graph is traversed. We posit that this is
expected behavior, necessary to continue ensuring that update times will remain linear. That techniques to reduce memory usage in DD are warranted.

The most surprising result is OWL2RL, in which both \textbf{Chibi} and \textbf{Relational}'s peak memory usage is 10x higher than \textbf{Diff}'s. This is explained by
the fact that \textbf{Diff} shares as much data as it is reasonable between rule executions. The other reasoners on the other hand, when executing the almost 100 rules
independently of each other, will have all of their rule execution indexes and data be in memory possibly at the same time, hence having a much larger memory footprint.
\section{Conclusion}
In this article we explored implementing an efficient datalog engine that uniformly handles additions and deletions with DD. We strayed from the state of the art by
not translating a datalog program to relational algebra, but in directly interpreting it. In order to give precise semantics about our proposal, we outlined it
as a non-incremental DBSP circuit, with the premise being that implementing it with DD's rust library will yield its incremental version.

Compared to the other reasoners that handled updates through delete-rederive, or with rematerialization in the case of Souffle, results showed that $\textbf{Diff}$ was
the only reasoner to exhibit uniform performance across updates, while using less memory than traditional methods. Relative to Souffle however, it used in the best case
2x more memory, and up to 500x more in the worst case. This highlighted the cost of DD, proved itself to being prohibitive in highly iterative scenarios.

We conclude that our contributions provide further evidence that DD is a promising platform for datalog.

This work could be expanded in multiple directions, as in supporting expressive variants of datalog, and in making it distributed, which is already supposed by DD's
underlying distribution logic, Timely Dataflow\cite{timely}.

\bibliographystyle{ACM-Reference-Format}
\bibliography{software}


\begin{thebibliography}{25}


\ifx \showCODEN    \undefined \def \showCODEN     #1{\unskip}     \fi
\ifx \showISBNx    \undefined \def \showISBNx     #1{\unskip}     \fi
\ifx \showISBNxiii \undefined \def \showISBNxiii  #1{\unskip}     \fi
\ifx \showISSN     \undefined \def \showISSN      #1{\unskip}     \fi
\ifx \showLCCN     \undefined \def \showLCCN      #1{\unskip}     \fi
\ifx \shownote     \undefined \def \shownote      #1{#1}          \fi
\ifx \showarticletitle \undefined \def \showarticletitle #1{#1}   \fi
\ifx \showURL      \undefined \def \showURL       {\relax}        \fi
\providecommand\bibfield[2]{#2}
\providecommand\bibinfo[2]{#2}
\providecommand\natexlab[1]{#1}
\providecommand\showeprint[2][]{arXiv:#2}

\bibitem[rus(t 08)]%
        {rust_ddflow}
 \bibinfo{year}{2023 [cited 2023 Oct 08]}\natexlab{}.
\newblock \bibinfo{title}{Differential Dataflow Implementation Github
  Repository [Internet]}.
\newblock
\urldef\tempurl%
\url{https://github.com/TimelyDataflow/differential-dataflow}
\showURL{%
\tempurl}


\bibitem[Abadi et~al\mbox{.}(2015)]%
        {differential_dataflow}
\bibfield{author}{\bibinfo{person}{Martín Abadi}, \bibinfo{person}{Frank
  McSherry}, {and} \bibinfo{person}{Gordon Plotkin}.}
  \bibinfo{year}{2015}\natexlab{}.
\newblock \showarticletitle{Foundations of Differential Dataflow}.
  \bibinfo{pages}{71--83}.
\newblock
\href{https://doi.org/10.1007/978-3-662-46678-0_5}{doi:\nolinkurl{10.1007/978-3-662-46678-0_5}}


\bibitem[Allemang and Hendler(2011)]%
        {rdfs}
\bibfield{author}{\bibinfo{person}{Dean~T. Allemang} {and}
  \bibinfo{person}{James~A. Hendler}.} \bibinfo{year}{2011}\natexlab{}.
\newblock \showarticletitle{Semantic Web for the Working Ontologist - Effective
  Modeling in RDFS and OWL, Second Edition}.
\newblock


\bibitem[Bader and Madduri(2006)]%
        {gtgraph}
\bibfield{author}{\bibinfo{person}{David~A. Bader} {and}
  \bibinfo{person}{Kamesh Madduri}.} \bibinfo{year}{2006}\natexlab{}.
\newblock \showarticletitle{GTgraph : A Synthetic Graph Generator Suite}.
\newblock


\bibitem[Budiu et~al\mbox{.}(2022)]%
        {dbsp}
\bibfield{author}{\bibinfo{person}{Mihai Budiu}, \bibinfo{person}{Frank
  McSherry}, \bibinfo{person}{Leonid Ryzhyk}, {and} \bibinfo{person}{Val
  Tannen}.} \bibinfo{year}{2022}\natexlab{}.
\newblock \showarticletitle{DBSP: Automatic Incremental View Maintenance for
  Rich Query Languages}.
\newblock \bibinfo{journal}{\emph{Proc. VLDB Endow.}}  \bibinfo{volume}{16}
  (\bibinfo{year}{2022}), \bibinfo{pages}{1601--1614}.
\newblock


\bibitem[Butvinik(2020)]%
        {online_machine_learning}
\bibfield{author}{\bibinfo{person}{Daniel~S. Butvinik}.}
  \bibinfo{year}{2020}\natexlab{}.
\newblock \showarticletitle{Online Machine Learning: An Introduction}.
\newblock
\urldef\tempurl%
\url{https://api.semanticscholar.org/CorpusID:235261920}
\showURL{%
\tempurl}


\bibitem[Cao et~al\mbox{.}(2013)]%
        {owl2rl}
\bibfield{author}{\bibinfo{person}{Son~Thanh Cao}, \bibinfo{person}{Linh~Anh
  Nguyen}, {and} \bibinfo{person}{Andrzej Szałas}.}
  \bibinfo{year}{2013}\natexlab{}.
\newblock \showarticletitle{The Web Ontology Rule Language OWL 2 RL + and Its
  Extensions}.
\newblock \bibinfo{journal}{\emph{Trans. Comput. Collect. Intell.}}
  \bibinfo{volume}{13} (\bibinfo{year}{2013}), \bibinfo{pages}{152--175}.
\newblock
\urldef\tempurl%
\url{https://api.semanticscholar.org/CorpusID:30277887}
\showURL{%
\tempurl}


\bibitem[Ceri et~al\mbox{.}(1989)]%
        {datalog}
\bibfield{author}{\bibinfo{person}{Stefano Ceri}, \bibinfo{person}{Georg
  Gottlob}, {and} \bibinfo{person}{Letizia Tanca}.}
  \bibinfo{year}{1989}\natexlab{}.
\newblock \showarticletitle{What you Always Wanted to Know About Datalog (And
  Never Dared to Ask).}
\newblock \bibinfo{journal}{\emph{Knowledge and Data Engineering, IEEE
  Transactions on}}  \bibinfo{volume}{1} (\bibinfo{date}{04}
  \bibinfo{year}{1989}), \bibinfo{pages}{146 -- 166}.
\newblock
\href{https://doi.org/10.1109/69.43410}{doi:\nolinkurl{10.1109/69.43410}}


\bibitem[Grosof et~al\mbox{.}(2003)]%
        {descr_to_datalog}
\bibfield{author}{\bibinfo{person}{Benjamin~N. Grosof}, \bibinfo{person}{Ian
  Horrocks}, \bibinfo{person}{Raphael Volz}, {and} \bibinfo{person}{S.
  Decker}.} \bibinfo{year}{2003}\natexlab{}.
\newblock \showarticletitle{Description logic programs: combining logic
  programs with description logic}. In \bibinfo{booktitle}{\emph{The Web
  Conference}}.
\newblock


\bibitem[Gupta and Mumick(1999)]%
        {dred}
\bibfield{author}{\bibinfo{person}{Ashish~Kumar Gupta} {and}
  \bibinfo{person}{Inderpal~Singh Mumick}.} \bibinfo{year}{1999}\natexlab{}.
\newblock \showarticletitle{Incremental Maintenance of Recursive Views: A
  Survey}.
\newblock


\bibitem[Imran et~al\mbox{.}(2022)]%
        {nexus}
\bibfield{author}{\bibinfo{person}{Muhammad Imran},
  \bibinfo{person}{G{\'a}bor~E. G{\'e}vay}, \bibinfo{person}{Jorge-Arnulfo
  Quian{\'e}-Ruiz}, {and} \bibinfo{person}{Volker Markl}.}
  \bibinfo{year}{2022}\natexlab{}.
\newblock \showarticletitle{Fast datalog evaluation for batch and stream graph
  processing}.
\newblock \bibinfo{journal}{\emph{World Wide Web}}  \bibinfo{volume}{25}
  (\bibinfo{year}{2022}), \bibinfo{pages}{971--1003}.
\newblock


\bibitem[Imran et~al\mbox{.}(2020)]%
        {cog}
\bibfield{author}{\bibinfo{person}{Muhammad Imran}, \bibinfo{person}{Gábor
  Gévay}, {and} \bibinfo{person}{Volker Markl}.}
  \bibinfo{year}{2020}\natexlab{}.
\newblock \bibinfo{booktitle}{\emph{Distributed Graph Analytics with Datalog
  Queries in Flink}}.
\newblock \bibinfo{pages}{70--83}.
\newblock
\showISBNx{978-3-030-61132-3}
\href{https://doi.org/10.1007/978-3-030-61133-0_6}{doi:\nolinkurl{10.1007/978-3-030-61133-0_6}}


\bibitem[Kotowski et~al\mbox{.}(2011)]%
        {ivm}
\bibfield{author}{\bibinfo{person}{Jakub Kotowski}, \bibinfo{person}{François
  Bry}, {and} \bibinfo{person}{Simon Brodt}.} \bibinfo{year}{2011}\natexlab{}.
\newblock \showarticletitle{Reasoning as Axioms Change - Incremental View
  Maintenance Reconsidered}. In \bibinfo{booktitle}{\emph{International
  Conference on Web Reasoning and Rule Systems}}.
\newblock
\urldef\tempurl%
\url{https://api.semanticscholar.org/CorpusID:17165711}
\showURL{%
\tempurl}


\bibitem[Motik et~al\mbox{.}(2015)]%
        {dredbf}
\bibfield{author}{\bibinfo{person}{Boris Motik}, \bibinfo{person}{Yavor Nenov},
  \bibinfo{person}{Robert Piro}, {and} \bibinfo{person}{Ian Horrocks}.}
  \bibinfo{year}{2015}\natexlab{}.
\newblock \showarticletitle{Incremental Update of Datalog Materialisation: the
  Backward/Forward Algorithm}. In \bibinfo{booktitle}{\emph{AAAI Conference on
  Artificial Intelligence}}.
\newblock


\bibitem[Motik et~al\mbox{.}(2019)]%
        {maintrevis}
\bibfield{author}{\bibinfo{person}{Boris Motik}, \bibinfo{person}{Yavor Nenov},
  \bibinfo{person}{Robert Piro}, {and} \bibinfo{person}{Ian Horrocks}.}
  \bibinfo{year}{2019}\natexlab{}.
\newblock \showarticletitle{Maintenance of datalog materialisations revisited}.
\newblock \bibinfo{journal}{\emph{Artif. Intell.}}  \bibinfo{volume}{269}
  (\bibinfo{year}{2019}), \bibinfo{pages}{76--136}.
\newblock
\urldef\tempurl%
\url{https://api.semanticscholar.org/CorpusID:62970849}
\showURL{%
\tempurl}


\bibitem[Mu{\~n}oz et~al\mbox{.}(2009)]%
        {rhodf}
\bibfield{author}{\bibinfo{person}{Sergio Mu{\~n}oz}, \bibinfo{person}{Jorge
  P{\'e}rez}, {and} \bibinfo{person}{Claudio Guti{\'e}rrez}.}
  \bibinfo{year}{2009}\natexlab{}.
\newblock \showarticletitle{Simple and Efficient Minimal RDFS}.
\newblock \bibinfo{journal}{\emph{J. Web Semant.}}  \bibinfo{volume}{7}
  (\bibinfo{year}{2009}), \bibinfo{pages}{220--234}.
\newblock
\urldef\tempurl%
\url{https://api.semanticscholar.org/CorpusID:9460781}
\showURL{%
\tempurl}


\bibitem[Murray et~al\mbox{.}(2013)]%
        {timely}
\bibfield{author}{\bibinfo{person}{Derek~Gordon Murray}, \bibinfo{person}{Frank
  McSherry}, \bibinfo{person}{Rebecca Isaacs}, \bibinfo{person}{Michael Isard},
  \bibinfo{person}{Paul Barham}, {and} \bibinfo{person}{Mart{\'i}n Abadi}.}
  \bibinfo{year}{2013}\natexlab{}.
\newblock \showarticletitle{Naiad: a timely dataflow system}.
\newblock \bibinfo{journal}{\emph{Proceedings of the Twenty-Fourth ACM
  Symposium on Operating Systems Principles}} (\bibinfo{year}{2013}).
\newblock


\bibitem[Poniszewska-Marańda and Czechowska(2021)]%
        {kubernetes}
\bibfield{author}{\bibinfo{person}{Aneta Poniszewska-Marańda} {and}
  \bibinfo{person}{Ewa Czechowska}.} \bibinfo{year}{2021}\natexlab{}.
\newblock \showarticletitle{Kubernetes Cluster for Automating Software
  Production Environment}.
\newblock \bibinfo{journal}{\emph{Sensors}} \bibinfo{volume}{21},
  \bibinfo{number}{5} (\bibinfo{year}{2021}).
\newblock
\showISSN{1424-8220}
\href{https://doi.org/10.3390/s21051910}{doi:\nolinkurl{10.3390/s21051910}}


\bibitem[Ryzhyk and Budiu(2019)]%
        {ddlog}
\bibfield{author}{\bibinfo{person}{Leonid Ryzhyk} {and} \bibinfo{person}{Mihai
  Budiu}.} \bibinfo{year}{2019}\natexlab{}.
\newblock \showarticletitle{Differential Datalog}. In
  \bibinfo{booktitle}{\emph{Datalog}}.
\newblock


\bibitem[Sandall(2023a)]%
        {rego}
\bibfield{author}{\bibinfo{person}{Torin Sandall}.}
  \bibinfo{year}{2023}\natexlab{a}.
\newblock
  \bibinfo{howpublished}{\url{https://www.openpolicyagent.org/docs/latest/policy-language/}}.
\newblock
\newblock
\shownote{Accessed: 2023-22-09}.


\bibitem[Sandall(2023b)]%
        {opa}
\bibfield{author}{\bibinfo{person}{Torin Sandall}.}
  \bibinfo{year}{2023}\natexlab{b}.
\newblock
  \bibinfo{howpublished}{\url{https://github.com/open-policy-agent/opa}}.
\newblock
\newblock
\shownote{Accessed: 2022-11-04}.


\bibitem[Scholz et~al\mbox{.}(2016)]%
        {souffle}
\bibfield{author}{\bibinfo{person}{Bernhard Scholz}, \bibinfo{person}{Herbert
  Jordan}, \bibinfo{person}{Pavle Subotic}, {and} \bibinfo{person}{Till
  Westmann}.} \bibinfo{year}{2016}\natexlab{}.
\newblock \showarticletitle{On fast large-scale program analysis in Datalog}.
\newblock \bibinfo{journal}{\emph{Proceedings of the 25th International
  Conference on Compiler Construction}} (\bibinfo{year}{2016}).
\newblock


\bibitem[Shkapsky et~al\mbox{.}(2016)]%
        {bigdatalog}
\bibfield{author}{\bibinfo{person}{Alexander Shkapsky}, \bibinfo{person}{Mohan
  Yang}, \bibinfo{person}{Matteo Interlandi}, \bibinfo{person}{Hsuan Chiu},
  \bibinfo{person}{Tyson Condie}, {and} \bibinfo{person}{Carlo Zaniolo}.}
  \bibinfo{year}{2016}\natexlab{}.
\newblock \showarticletitle{Big Data Analytics with Datalog Queries on Spark}.
\newblock \bibinfo{journal}{\emph{Proceedings. ACM-Sigmod International
  Conference on Management of Data}}  \bibinfo{volume}{2016},
  \bibinfo{pages}{1135--1149}.
\newblock
\href{https://doi.org/10.1145/2882903.2915229}{doi:\nolinkurl{10.1145/2882903.2915229}}


\bibitem[Szab{\'o} et~al\mbox{.}(2021)]%
        {laddder}
\bibfield{author}{\bibinfo{person}{Tam{\'a}s Szab{\'o}},
  \bibinfo{person}{Sebastian Erdweg}, {and} \bibinfo{person}{G{\'a}bor
  Bergmann}.} \bibinfo{year}{2021}\natexlab{}.
\newblock \showarticletitle{Incremental whole-program analysis in Datalog with
  lattices}.
\newblock \bibinfo{journal}{\emph{Proceedings of the 42nd ACM SIGPLAN
  International Conference on Programming Language Design and Implementation}}
  (\bibinfo{year}{2021}).
\newblock
\urldef\tempurl%
\url{https://api.semanticscholar.org/CorpusID:235474488}
\showURL{%
\tempurl}


\bibitem[Zhu et~al\mbox{.}(2018)]%
        {recstep}
\bibfield{author}{\bibinfo{person}{Jianqiao Zhu}, \bibinfo{person}{Zuyu Zhang},
  \bibinfo{person}{Aws Albarghouthi}, \bibinfo{person}{Paraschos Koutris},
  {and} \bibinfo{person}{Jignesh Patel}.} \bibinfo{year}{2018}\natexlab{}.
\newblock \showarticletitle{Scaling-Up In-Memory Datalog Processing:
  Observations and Techniques}.
\newblock  (\bibinfo{date}{12} \bibinfo{year}{2018}).
\newblock


\end{thebibliography}

\end{document}